\documentclass[12pt]{article}

\usepackage{graphics,epsfig}
\usepackage[latin1]{inputenc}
\usepackage[english,activeacute]{babel}
\usepackage[]{graphicx}
\usepackage{latexsym}
\usepackage{amsmath, amsthm, amsfonts, amssymb}
\usepackage{verbatim}
\usepackage[all]{xy}
\usepackage{color}
\usepackage{amssymb}

\begin{document}

\newtheorem{theo}{Theorem}[section]
\newtheorem{definition}[theo]{Definition}
\newtheorem{lem}[theo]{Lemma}
\newtheorem{prop}[theo]{Proposition}
\newtheorem{coro}[theo]{Corollary}
\newtheorem{exam}[theo]{Example}
\newtheorem{rema}[theo]{Remark}
\newtheorem{example}[theo]{Example}
\newtheorem{principle}[theo]{Principle}
\newcommand{\ninv}{\mathord{\sim}}
\newtheorem{axiom}[theo]{Axiom}
\title{Solutions for the MaxEnt problem with symmetry constraints}

\author{Marcelo Losada$^{1}$, Federico Holik$^{2}$, Cesar Massri$^{3}$ and Angelo Plastino$^{2}$}
\maketitle
\begin{center}

\begin{small}
1- Universidad de Buenos Aires - CONICET, Ciudad de Buenos Aires, Argentina \\
2- Department of Mathematics, University CAECE - CONICET IMAS\\
3- National University La Plata - CONICET IFLP-CCT, C.C. 727 - 1900
La Plata, Argentina\\
\end{small}

\end{center}

\begin{abstract}
In this paper, we deal with the situation in which the unknown state
of a quantum system has to be estimated under the assumption that it
is prepared obeying a known set of symmetries. We present a system
of equations and an explicit solution for the problem of determining
the MaxEnt state satisfying these constraints. Our approach can be
applied to very general situations, including symmetries of the
source represented by Lie and finite groups.
\end{abstract}

\section{Introduction}

\noindent The principle of maximum entropy (MaxEnt principle) is a
powerful technique for estimating states of probabilistic models
\cite{Jaynes-1957a,Jaynes-1957b,katz,Presse-MaxEntMaxCalRevModPhys}.
This principle states that the most suitable probability
distribution compatible with the known data is the one with largest
entropy \cite{Jaynes-1957a,Jaynes-1957b}. It finds applications in
many fields of research
\cite{Presse-MaxEntMaxCalRevModPhys,Cavagna-2014-BirdsMaxEnt,BerettaMaxEnt-2014,Sinatra-MaxEntRandomWalks-2011,Diambra-Plastino-PRE-MaxEnt-1995,Rebollo-Neira-Plastino-PRE-MaxEnt-2001,Diambra-Plastino-PRE-MaxEnt-1996,Goswami-Prasad-PRD-MaxEnt-2013,Cofre-Cessac-MaexEntNeuralNetworks,Lanata-MaximumEntanglement-2014,Canosa-Plastino-Rossignoli-PRA-MaxEnt,Plastino-Portesi-PRA-MaxEnt,Tkacic-MaxEntNeuralNetwork-2013,Goncalves-2013-MaxEntTomography,Gzyl,Alves-2015,Plastino-2007}.
In particular, it has become very useful in quantum information
theory for estimating quantum states
\cite{Goncalves-2013-MaxEntTomography,MaxEnt-Tomography,Ziman}. It
can also be extended to a very general family of probabilistic
models \cite{Holik-Plastino-QuantalEffects-MaxEnt}.

In this paper, we want to apply the MaxEnt principle to physical
systems with symmetry constraints. Symmetries play a key role in the
study of physical systems. They are mathematically described by
group transformations that leave some physical properties unchanged.
The problem of determining the MaxEnt state for generalized
probabilistic models under quite general symmetry constraints was
studied in \cite{Holik-Massri-Plastino-IJGMMP-2016}. In that paper,
the MaxEnt approach was reformulated allowing the inclusion of group
transformations representing physical symmetries, and the conditions
for the existence of the solution were determined.

\vskip 3mm

In this work, we take a step further by presenting an analytical
expression for the MaxEnt state with symmetry constraints, suitable
for a vast family of quantum models. This can be useful for the
problem of quantum state estimation in quantum information
processing problems. As is well known, the number of measurements
needed to perform a complete quantum tomography on an unknown state,
increases with the dimension of the system \cite{QSE-Paris,LR1}.
Thus, this procedure becomes inviable in practice, specially for
multi-qubit states, for which the number of measurements needed
scales exponentially with the number of particles. This situation
gives a major relevance to the quantum state estimation techniques
that use fewer measurements. In this paper we deal with the
situation in which the state estimation using MaxEnt is performed
under the assumption that the state is prepared obeying a known set
of symmetries. This assumption, reduces the complexity of the
problem in a substantial way
\cite{Holik-Massri-Plastino-IJGMMP-2016}. Our approach can be
applied to very general situations, including symmetries of the
source represented by both, Lie and finite groups.

The paper is organized as follows. In Section~\ref{s:Formulation},
we revise the traditional version of the MaxEnt estimation problem,
and we reformulate it by including additional symmetry constraints.
In Section \ref{s:Motivation}, we discuss a classical example that
shows the need of a systematic treatment of the problem. Moreover,
we consider the general case of the classical MaxEnt estimation
problem for a finite system with symmetry constraints. In Section
\ref{s:QuantumProblem}, we present the solution to the quantum
MaxEnt problem with symmetries constraints. Finally, in Section
\ref{s:Conclusions}, we draw some conclusions.

\section{Principle of maximum entropy}\label{s:Formulation}

The MaxEnt principle states that the probability distribution which
best fits the available information is the one that maximizes the
entropy (also called missing information). This principle was first
introduced by E. Jaynes in two seminal papers
\cite{Jaynes-1957a,Jaynes-1957b}, where he emphasized a natural
correspondence between statistical mechanics and information theory.
He argued that classical and quantum statistical mechanics can be
formulated on the basis of information theory if the probability
distribution or the density operator are obtained from the MaxEnt
principle \cite{Jaynes-1957a,Jaynes-1957b,katz}.

In classical mechanics the probability distribution is defined over
the phase space of the system and its entropy is given by Shannon's
entropy. In quantum mechanics, states are described by density
operators acting on a Hilbert space and the entropy of the system is
given by von Neumann's entropy. Von Neumann's entropy can be
considered as a natural non-commutative generalization of Shannon's
entropy \cite{Holik-Entropy,Holik-Plastino-Saenz}. While in this
work we use the Shannon and von Neumann entropies, it is important
to remark that there exist other entropic quantities that find many
applications in diverse fields of research
\cite{Holik-Entropy,Renyi,Tsallis,Holik-Bosyk-qinp,Holik-Portesi-2018,Rastegin-2013,Zhang-2015,Tanak-2017,Kurzyk-2018,Liang}.
In the case that the available information is given by the mean
value of some set of observables, the probability distribution can
be obtained using Lagrange multipliers. In the classical case, these
observables are functions over phase space, and in the quantum case
they are self-adjoint operators acting on a Hilbert space.

The classical and quantum versions of the MaxEnt problem can be
stated as follows\footnote{We restrict ourselves to  finite cases in
order to simplify the exposition. For a more general treatment see
\cite{katz}.}:
\begin{itemize}
\item {\bf Classical MaxEnt}

Given $n$ observables $A_i$ ($1\leq i \leq n$), with $m$ outcomes
$A_{i}(j)$ ($1\leq j \leq m$), determine the probability
distribution $p_j$ which maximizes the Shannon entropy
\begin{equation}
H_{S} = - \sum_{j=1}^n  p_{j} \ln p_j,
\end{equation}
and satisfies the constraints
\begin{equation}
\langle A_{i}\rangle = \sum_{j=1}^m p_j  A_i(j)  = a_i, ~~~ \forall
\,  i= 1, \ldots, n.
\end{equation}

\vspace{5px}

The solution is given by (see for example \cite{Jaynes-1957a})
\begin{equation}\label{classical maxent}
p_j= \frac{e^ {\sum_{i=1}^n \lambda_{i} A_{i}(j)}}{Z}, ~~~   1\leq j
\leq m.
\end{equation}
where $Z = \sum_{j=1}^m e^ {\sum_{i=1}^n \lambda_{i} A_{i}(j)}$ is
the partition function, and the Lagrange multipliers $\lambda_{i}$
are given by the relation
\begin{equation}\label{classical lagrange multipliers}
a_{i}= \frac{\partial}{\partial\lambda_{i}}\ln Z,\quad    1\leq i
\leq n.
\end{equation}

\item {\bf Quantum MaxEnt}

Given $n$ observables $\hat{A}_i$ ($1\leq i \leq n$), determine the
density matrix $\hat{\rho}$ which maximizes the von Neumann entropy
\begin{equation}
H_{VN} = - \mbox{Tr} (\hat{\rho} \ln\hat{\rho}),
\end{equation}
and satisfies the constraints
\begin{equation}
\langle \hat{A}_{i}\rangle=  \mbox{Tr} (\hat{\rho} \hat{A}_i)  =
a_i, ~~~ \forall \,  i= 1, \ldots, n.
\end{equation}

\vspace{5px}

The solution is given by (see for example \cite{Jaynes-1957b})
\begin{equation}
\hat{\rho} = \frac{e^ {\sum_{i=1}^n \lambda_{i}\hat{A}_{i}}}{Z},
\end{equation}
with $Z = \mbox{Tr}(e^{\sum_{i=1}^n \lambda_{i} \hat{A}_{i} })$, and
the Lagrange multipliers are given by the relations
\begin{equation}
a_{i}= \frac{\partial}{\partial\lambda_{i}}\ln Z,\quad  1\leq i \leq
n.
\end{equation}
\end{itemize}

In this work, we are going to consider the MaxEnt problem with
additional constraints given by symmetries represented by the action
of a group $\mathcal{G}$. More specifically, we aim to solve the
following problems:

\begin{itemize}
\item {\bf Classical MaxEnt with symmetries} 

Given a group $\mathcal{G}$ and $n$ observables $A_i$ ($1\leq i \leq
n$), with $m$ outcomes $A_{i}(j)$ ($1\leq j \leq m$), determine the
probability distribution $p_j$ (i.e., $\sum_{j=1}^{m}p_j =1$ and
$p_j \geq 0$ for $1\leq j \leq m$) which maximizes the Shannon
entropy and satisfies the constraints
\begin{align}
\langle A_{i}\rangle &= \sum_{j=1}^m p_j  A_i(j)  = a_i, ~~~ \forall \,  i= 1, \ldots, n.\\
p_{g (j)} &=  p_j, ~~~ \forall \, g \in \mathcal{G}, ~~~ \forall \,
j= 1, \ldots, m.
\end{align}

\item {\bf Quantum MaxEnt with symmetries}

Given a group $\mathcal{G}$ representing a physical symmetry and $n$
observables $\hat{A}_i$ ($1\leq i \leq n$), determine the density
matrix $\hat{\rho}$ which maximizes the von Neumann entropy, and
satisfies the constraints
\begin{align}
\langle \hat{A}_{i}\rangle &=  \mbox{Tr} (\hat{\rho} \hat{A}_i)  = a_i, ~~~ \forall \,  i= 1, \ldots, n. \\
\hat{U}_{g} \hat{\rho} \hat{U}_{g}^{\dag}&= \hat{\rho}, ~~~ \forall
\, g \in \mathcal{G},
\end{align}

where $\hat{U}_{g}$ is the unitary representation of the group
element $g$.
\end{itemize}

The conditions for the existence of the solutions of these two
problems (under more general constraints) were discussed in
\cite{Holik-Massri-Plastino-IJGMMP-2016}. In the following sections,
we look for solutions to the classical and quantum MaxEnt problems
with symmetry constraints given by a group $\mathcal{G}$.

\section{Motivation: Classical MaxEnt}\label{s:Motivation}

In this section we consider a classical and discrete example of the
MaxEnt problem with and without symmetry constraints, in order to
illustrate the differences between both cases. This simple example
serves as a motivation for the systematic treatment of the problem,
that we present for the quantum case in the next section. Moreover,
we consider the general case of the classical MaxEnt problem for a
finite system with symmetry constraints.

A fair dice must be manufactured in such a way that all faces are
equivalent -from a physical point of view and for all practical
purposes. In that case, there is a symmetry group that leaves
invariant the probabilities of all outcomes, equal to $1/6$. The
action of an element of this group is represented by a permutation
of the outcomes.

Now suppose that the dice is fabricated in such a way that the sixth
face is heavier than the others, and the remaining faces are
designed in a equivalent way. It is not hard to imagine an scenario
in which such a breaking of the symmetry makes the sixth face more
likely than the others, while faces $2$, $3$, $4$, and $5$ are
equally likely. Then, the symmetry group of the loaded dice is
reduced to a subgroup of the full symmetry group. This subgroup,
which we call $\mathcal{G}$, leave invariant the probabilities of
faces $2$, $3$, $4$, and $5$, and thus, it is formed by all possible
permutations of these outcomes. In what follows, we are going to
show that, without taking into account the constraints of the dice
given by the symmetry group $\mathcal{G}$, the MaxEnt solution for
the probabilities is not adequate.

The set of outcomes of the die is given by $\Lambda =
\{1,2,3,4,5,6\}$ and $p_1,\ldots,p_6$ are the corresponding
probabilities. We consider the observable $A$ taking the values
$A(j) = j$ for all $j\in \Lambda$. The constraint given by the mean
value $\langle A\rangle$ of observable $A$ is $\sum_{j=1}^{6} j p_j
= a$, where $a$ is a real number between $1$ and $6$. A fair dice
satisfies $\langle A\rangle=3.5$.

On the one hand, we consider the situation in which we only know
that the loaded dice has mean value $\langle A\rangle=a$. If we
consider the solution of the MaxEnt problem without taking into
account the symmetry constraints (i.e., without considering that the
faces $2$, $3$, $4$, and $5$ are all equally likely), according to
equation \eqref{classical maxent}, we obtain the following
probabilities
\begin{equation}
p_j = \frac{e ^ {j\lambda}}{Z},\quad  1\leq j \leq m,
\end{equation}
with $Z = \sum_{j=1}^6  e ^ {j \lambda}$, and the Lagrange
multiplier $\lambda$ is given by equation \eqref{classical lagrange
multipliers},
\begin{equation}
a = \frac{\sum_{j=1}^{6} j e^{j \lambda}}{Z}.
\end{equation}
It should be noted that for $a \neq 3.5$, we obtain $\lambda \neq
0$, and therefore, all the $p_{j}$'s are different from each other.
This solution does not satisfy the symmetry constraint.

On the other hand, we consider that we know that there is a symmetry
in the fabrication process of the dice implying that $p_2= p_3 =
p_4= p_5$. This symmetry is described by the group $\mathcal{G}$
generated by all possible transformations which permute outcomes
$2,3,4,5$. Under these conditions, the MaxEnt problem consists of
obtaining the probability distribution $p_1$, $p_2= p_3= p_4= p_5$,
$p_6$ which maximizes the Shannon entropy
\begin{equation}
H_{S} = -  p_{1} \ln p_1 - 4  p_{2} \ln p_2 - p_{6} \ln p_6,
\end{equation}
and satisfies the constraints
\begin{equation}
\sum_{j=1}^{6}p_j= p_{1} + 4 p_{2} + p_{6} = 1, ~~~~~  \langle A
\rangle = a  = p_1  + 14 p_2 + 6 p_6.
\end{equation}
To solve this problem we can use the method of Lagrange multipliers.
We define the Lagrangian function
\begin{equation}
\mathcal{L}(p_1, p_2, p_6) = -  p_{1} \ln p_1 - 4  p_{2} \ln p_2 -
p_{6} \ln p_6 + \lambda_0 ( p_{1} + 4 p_{2} + p_{6} - 1)+ \lambda (
p_1  + 14 p_2 + 6 p_6 - a), \nonumber
\end{equation}
and we find the stationary points, i.e., we solve the following
equations
\begin{align}
\frac{\partial \mathcal{L}}{\partial p_1} &= -  \ln p_1 - 1  + \lambda_0 + \lambda = 0,\\
\frac{\partial \mathcal{L}}{\partial p_2} &=-  4\ln p_2 - 4  + 4 \lambda_0 + 14 \lambda = 0, \\
\frac{\partial \mathcal{L}}{\partial p_6} &= -  \ln p_6 - 1  +
\lambda_0 + 6 \lambda = 0.
\end{align}
It easy to see that the probability distribution is given by
\begin{equation}
p_1 = \frac{e ^ { \lambda}}{Z} , \quad p_j = \frac{e ^ {3.5
\lambda}}{Z} \quad j= 2,\ldots,5, \quad p_6 = \frac{e ^ {6
\lambda}}{Z},
\end{equation}
with $Z = e ^ {1 - \lambda_0}= e ^ { \lambda}+4 e ^ {3.5 \lambda} +e
^ {6 \lambda}$, and $\lambda$ is obtained from the relation $a =
\frac{\partial}{\partial \lambda} \ln Z$
\begin{equation}
a = \frac{e ^ { \lambda}+14 e ^ {3.5 \lambda} +6 e ^ {6
\lambda}}{Z}.
\end{equation}

In this way, we arrive at a situation in which the solution of the
MaxEnt problem can be  different if we introduce information
concerning its symmetries.

If we want to deal with more complex problems, a systematic
treatment of the problem is needed. In what follows, we discuss the
general case of the classical MaxEnt problem with symmetry
constraints.

We consider a classical physical system with a finite set of states
given by $\Lambda = \{1, \ldots , m \}$, with $p_1, \ldots, p_m$
being the corresponding probabilities. Moreover, we consider $n$
observables $A_i$,  with $m$ outcomes $A_{i}(j)$ ($1\leq j \leq m$),
and mean values $\langle A_i \rangle = \sum_{j=1}^m p_j  A_i(j)=
a_i$. The system also has a symmetry constraint given by a group
$\mathcal{G}$, which implies
\begin{equation}
p_j = p_{g(j)}, ~~~ \forall \, g \in \mathcal{G}, ~~~ \forall \,  j=
1, \ldots, m.
\end{equation}

The MaxEnt principle implies maximizing Shannon entropy
\begin{equation}
H_{S} = - \sum_{j=1}^n  p_{j} \ln p_j,
\end{equation}
under the following constraints
\begin{align}
p_j &= p_{g(j)}, ~~~ \forall \, g \in \mathcal{G}, ~~~~~~ \forall \,  j= 1, \ldots, m,  \label{symmetry constraints} \\
1 &= \sum_{j=1}^m p_j, \label{normalization constraints}\\
\langle A_{i}\rangle &= \sum_{j=1}^m p_j  A_i(j)  = a_i, ~~~~~~
\forall \,  i= 1, \ldots, n. \label{mean value constraints}
\end{align}

The symmetries constraints \eqref{symmetry constraints} allow to
regroup the probabilities $p_j$ in sets of equal probabilities. That
means that there are $r$ sets of indexes $\mathcal{J}_l \subseteq
\{1, \ldots, m \}$, with $ 1\leq l \leq r$, such that
\begin{equation}
\bigcup_{l=1}^r \mathcal{J}_l = \{1, \ldots,m\}, \qquad
\mathcal{J}_l \cap \mathcal{J}_{l'} = \emptyset,  ~~ \,l\neq l'
\end{equation}
and $p_j= p_{j'}$, if and only if, there is a set $\mathcal{J}_l$
such that $j, j' \in \mathcal{J}_l$.

Let $d_l$ be the cardinality of $\mathcal{J}_l$, and for each $l=
1,\ldots,r$, we define $q_l = p_j$, with $j$ an arbitrary index in
$\mathcal{J}_l$. Now, we can express the Shannon entropy and the
constraint equations \eqref{symmetry constraints},
\eqref{normalization constraints}, \eqref{mean value constraints} in
terms of $q_l$,
\begin{align*}
H_{S} &= - \sum_{j=1}^m  p_{j} \ln p_j = - \sum_{l=1}^r d_l q_{l} \ln q_l  \\
1 &= \sum_{j=1}^m  p_{j} = \sum_{l=1}^r d_l q_{l}  \\
\langle A_i \rangle& = a_i = \sum_{j=1}^m p_j  A_i(j)= \sum_{l=1}^r
q_l \sum_{j\in \mathcal{J}_l}  A_i(j)= \sum_{l=1}^r d_l q_l
\tilde{A}_i(l), ~~~ \forall \,  i= 1, \ldots, n,
\end{align*}
with $\tilde{A}_i(l)= \frac{1}{d_l}\sum_{j\in \mathcal{J}_l}A_i(j)$.

To maximize the Shannon entropy, we use the method of Lagrange
multipliers. The Lagrangian function is given by
\begin{equation}
\mathcal{L}(q_1, ..., q_r) = - \sum_{l=1}^r d_l q_{l} \ln q_l +
\lambda_0 (\sum_{l=1}^r d_l q_{l} - 1)+ \sum_{i=1}^n \lambda_i
\left( \sum_{l=1}^r d_l q_l  \tilde{A}_i(l) - a_i \right).
\end{equation}
The equations for the stationary points are the following ones
\begin{equation}
\frac{\partial \mathcal{L}}{\partial q_l} = - d_l \ln q_l - d_l  +
d_l \lambda_0 + d_l \sum_{i=1}^n \lambda_i \tilde{A}_i(l)  = 0.
\end{equation}
It easy to see that the probability distribution is given by
\begin{equation}
q_l = \frac{e ^ {\sum_{i=1}^n \lambda_i \tilde{A}_{i}(l) }}{Z},
\quad\mbox{with} \quad Z = e^{1- \lambda_0}= \sum_{l=1}^r d_l e ^
{\sum_{i=1}^n \lambda_i \tilde{A}_{i}(l) },
\end{equation}
and the Lagrange multipliers $\lambda_i$ are obtained from the
equations $a_i = \frac{\partial}{\partial \lambda_i} \ln Z $,
\begin{equation}
a_i = \frac{\sum_{l=1}^r d_l  \tilde{A}_{i}(l) e^{\sum_{i=1}^n
\lambda_i \tilde{A}_{i}(l)}}{Z}.
\end{equation}

The classical MaxEnt problem illustrates the relevance of taking
into account the symmetry constraints for estimating probabilities.
In the next section we discuss the quantum case.

\section{Quantum MaxEnt with symmetries}\label{s:QuantumProblem}

In this section, we study the quantum MaxEnt problem with symmetry
constraints. We consider a quantum physical system with Hilbert
space $\mathcal{H}$ and $n$ observables $\hat{A}_i$. Their mean
values are given by $\mbox{Tr} (\hat{\rho} \hat{A}_i)=a_i$, with
$\hat{\rho}$ an unknown quantum state.

\subsection{Lie groups} \label{lie groups}

We start with of a system that has a symmetry given by a continuous
group $\mathcal{G}$. We will work with the unitary representation of
$\mathcal{G}$ in the Hilbert space and we assume that it is a
connected Lie group. Hence, any $g\in\mathcal{G}$ will be
represented by a unitary operator $\hat{U}_{g}$, and -as it is well
known- we can write $\hat{U}_{g}= e^{i\hat{Q}}$, where $\hat{Q}$ is
a self adjoint operator \cite[12.37]{rudin}.

We are looking for states $\hat{\rho}$ which are invariant under the
action of the symmetry group. Therefore, they have to satisfy the
condition
\begin{equation}
\hat{U}_{g} \hat{\rho} \hat{U}_{g}^{\dag}= \hat{\rho}, ~~~ \forall
\, g \in \mathcal{G}.
\end{equation}
Since the Lie group is connected, the above condition is also valid
for  elements of the form $U_{g}=e^{i\hat{Q}t}$, with $t$ a real
parameter belonging to some interval. By considering elements of
this form and using the Taylor expansion, we obtain:
\begin{equation}\label{symmetry condition}
\hat{U}_{g} \hat{\rho} \hat{U}_{g}^{\dag}-\hat{\rho} = e^{i t
\hat{Q}} \hat{\rho} e^{- i t \hat{Q}}- \hat{\rho} =  it[\hat{Q};
\hat{\rho}] + o(t^2) = 0,
\end{equation}
where $[\hat{Q};\hat{\rho}]$ is the commutator between $\hat{Q}$ and
$\hat{\rho}$. The condition \eqref{symmetry condition} is valid for
all values of the parameter $t$ and all elements $\hat{Q}$ of the
Lie algebra $\mathfrak{g}$ associated with $\mathcal{G}$. As the
polynomial functions in the expansion are linearly independent, this
condition can only be satisfied if
\begin{equation}\label{eq-lie}
[\hat{\rho};\hat{Q}]= 0,\quad \forall \hat{Q} \in \mathfrak{g}.
\end{equation}
If we consider a set of generators $\{\hat{Q}_k\}_{k\in I}$ ($I$ a
set of indexes) of the Lie algebra $\mathfrak{g}$, the condition
\eqref{eq-lie} can be expressed as follows
\begin{equation}\label{eq-lie2}
[\hat{\rho};\hat{Q}_k]= 0,\quad \forall \hat{Q}_k \, \in \,
\{\hat{Q}_k\}_{k\in I}.
\end{equation}

Therefore, the quantum MaxEnt problem with symmetry constraints can
be reformulated as follows:

\vspace{10px}
\begin{itemize}
\item \textbf{Quantum MaxEnt with symmetries:} \textit{Given a connected unitary Lie group $\mathcal{G}$, and $n$ observables $\hat{A}_i$ ($1\leq i \leq n$), determine the density matrix $\hat{\rho}$ which maximizes the von Neumann entropy}
\begin{equation}
H_{VN} = - \mbox{Tr} (\hat{\rho} \ln\hat{\rho}),
\end{equation}
\textit{and satisfies the constraints}
\begin{align}
\langle \hat{A}_{i}\rangle&=  \mbox{Tr} (\hat{\rho} \hat{A}_i)  = a_i, ~~~ \forall \,  i= 1, \cdots, n, \\
[\hat{\rho};\hat{Q}_k]&= 0, \quad \forall \hat{Q}_k  \in
\{\hat{Q}_k\}_{k\in I},
\end{align}
\textit{where $\{\hat{Q}_k\}_{k\in I}$ is a set of generators of the
Lie Algebra associated with $\mathcal{G}$.}
\end{itemize}

\vspace{10px}

In what follows, we give an explicit solution of this problem for
quantum systems represented by finite dimensional Hilbert spaces.
The infinite dimensional case is more complicated and will be
treated elsewhere.

Let $m$ be the dimension of the Hilbert space $\mathcal{H}$, and
$\{\hat{O_j}\}_{1\leq j\leq m^2}$ a basis for the space of linear
operators $\mathcal{L}(\mathcal{H})$. First, we note that
\begin{equation}
[\hat{ \rho}; \hat{Q}_k]= 0 ~ \Longleftrightarrow ~
\mbox{Tr}([\hat{\rho};\hat{Q}_k]\hat{O_j})=0, ~~~ \forall \,  j= 1,
\ldots, m^2.
\end{equation}
Then, using the cyclic property of the trace, we obtain
$\mbox{Tr}([\hat{\rho};\hat{Q}_k]\hat{O_j})=
\mbox{Tr}(\hat{\rho}[\hat{Q}_k;\hat{O_j}])$. Therefore, we conclude
\begin{equation}
[\hat{ \rho}; \hat{Q}_k]= 0 ~ \Longleftrightarrow ~
\mbox{Tr}(\hat{\rho}[\hat{Q}_k;\hat{O_j}])=0, ~~~ \forall \,  j= 1,
\ldots, m^2.
\end{equation}
In particular, it is possible to choose the basis
$\{\hat{O_j}\}_{1\leq j\leq m^2}$ in such a way that all the
operators $\hat{O_i}$ are Hermitian. Moreover, since the generators
of the Lie Algebra $\hat{Q}_k$ are Hermitian, the commutators
$[i\hat{Q}_k;\hat{O_j}]$ are also Hermitian. Then,
\begin{equation}\label{solucion 1}
[\hat{ \rho}; \hat{Q}_k]= 0 ~ \Longleftrightarrow ~
\mbox{Tr}(\hat{\rho}[i\hat{Q}_k;\hat{O_j}])=0, ~~~ \forall \,  j= 1,
\ldots, m^2,
\end{equation}
with $\{\hat{O_j}\}_{1\leq j\leq m^2}$ an Hermitian basis of
$\mathcal{L}(\mathcal{H})$. Since $[i\hat{Q}_k;\hat{O_j}]$ are
Hermitian operators, the conditions
$\mbox{Tr}(\hat{\rho}[i\hat{Q}_k;\hat{O_j}])=0$ can be considered as
constraints for the mean values of a family of auxiliary observables
$\{[i\hat{Q}_k;\hat{O_j}]\}_{k\in I;j= 1,\ldots,m^2}$. It should be
noted that we have proved that \emph{the symmetries constraints can
be rewritten as linear constraints}.

Therefore, for finite dimensional models, the quantum MaxEnt problem
with symmetry constraints is that of determining a density matrix
$\hat{\rho}$ which maximizes the von Neumann entropy, and satisfies
the constraints
\begin{align}
\langle \hat{A}_{i}\rangle&=  \mbox{Tr} (\hat{\rho} \hat{A}_i)  = a_i, ~~~ \forall \,  i= 1, \ldots, n,\\
\langle [i\hat{Q}_k;\hat{O_j}] \rangle & =
\mbox{Tr}(i\hat{\rho}[\hat{Q}_k;\hat{O_j}])=0, ~~~ \forall k \in I,
~ \forall \,  j= 1, \ldots, m^2.
\end{align}
The advantage of this formulation of the problem is that the
solution is straightforward. Since the extra conditions are also
mean values constraints, the solution has the same form that the
standard MaxEnt problem. The explicit solution is given by
\begin{equation}\label{general solution}
\hat{\rho} = \frac{e^ {\sum_{i=1}^n
\lambda_{i}\hat{A}_{i}+\sum_{k\in
I}\sum_{j=1}^{m^2}\gamma_{k,j}[i\hat{Q}_k;\hat{O_j}]}}{Z},
\end{equation}
where $\{\hat{Q}_k\}_{k\in I}$ is a set of generators of the Lie
algebra, $\{\hat{O_j}\}_{1\leq j\leq m^2}$ is an Hermitian basis of
$\mathcal{L}(\mathcal{H})$, and $Z=
\mbox{Tr}(e^{\sum_{i=1}^n\lambda_{i}\hat{A}_{i}+\sum_{k\in
I}\sum_{j=1}^{m^2}\gamma_{k,j}[i\hat{Q}_k;\hat{O_j}]})$. The
Lagrange multipliers satisfy the relations
\begin{align}
a_{i}&= \frac{\partial}{\partial\lambda_{i}}\ln Z,\quad  1\leq i \leq n, \label{lagrangian multiplier 1}\\
0 &= \frac{\partial}{\partial\gamma_{k,j}}\ln Z,\quad  1\leq j \leq
m^2, \, k \in I. \label{lagrangian multiplier 2}
\end{align}

In what follows, we illustrate with some examples how this method
works.

\subsection{Finite groups}

In this section, we focus on a system whose symmetries are
represented by a finite group $\mathcal{G}$. Again, we will work
with the unitary representation of $\mathcal{G}$ in the Hilbert
space $\mathcal{H}$. Since $\mathcal{G}$ is a finite group, each
element $\hat{U}\in \mathcal{G}$ can be written as
\begin{equation}
\hat{U}=\hat{U}_{k_{1}}\ldots \hat{U}_{k_{n}}
\end{equation}

\noindent where the $\hat{U}_{k}$ are the generators of the group,
with $k \in I$ and $I$ a set of indexes. The condition $\hat{U}
\hat{\rho} \hat{U}^\dagger=\hat{\rho}$ for all $\hat{U}\in G$
implies that $\hat{U}_{k}\hat{\rho} \hat{U}_{k}^\dagger=\hat{\rho}$
for all generators of the group. The converse is also true: if
$\hat{U}_{k}\hat{\rho} \hat{U}_{k}^\dagger=\hat{\rho}$ for all
generator $\hat{U}_k$, then, for an aritrary element $\hat{U}\in G$,
we have
\begin{equation}
\hat{U}\hat{\rho} \hat{U}^\dagger=\hat{U}_{k_{1}}\ldots
\hat{U}_{k_{n}}\hat{\rho} \hat{U}^\dagger_{k_{n}}\ldots
U_{k_{1}}^\dagger=\hat{U}_{k_{1}}\ldots \hat{U}_{k_{n-1}}\hat{\rho}
U^\dagger_{k_{n-1}}\ldots
\hat{U}_{k_{1}}^\dagger=\ldots=\hat{U}_{k_{1}}\hat{\rho}
U_{k_{1}}^\dagger=\hat{\rho}
\end{equation}
Therefore, the state is invariant under all elements of the group
if, and only if, it is invariant under the set of generators.

Then, proceeding similarly as in Section \ref{lie groups}, we choose
an hermitian basis  $\{\hat{O_j}\}_{1\leq j\leq m^2}$  for the space
of linear operators $\mathcal{L}(\mathcal{H})$ (with $m$  the
dimension of $\mathcal{H}$), and we note that
\begin{equation}
[\hat{ \rho}; \hat{U}_k]= 0 ~ \Longleftrightarrow ~
\mbox{Tr}(\hat{\rho}[i\hat{U}_k;\hat{O_j}])=0, ~~~ \forall \,  j= 1,
\ldots, m^2,
\end{equation}

Therefore, the quantum MaxEnt problem with a symmetry given by a
finite group is equivalent to determe a density matrix $\hat{\rho}$
which maximizes the von Neumann entropy, and satisfies the following
constraints
\begin{align}
\langle [i\hat{U}_k;\hat{O_j}] \rangle & =
\mbox{Tr}(i\hat{\rho}[\hat{U}_k;\hat{O_j}])=0, ~~~ \forall k \in I,
~ \forall \,  j= 1, \ldots, m^2.
\end{align}
These equations define the conditions that come from group
invariance. Extra conditions must be added if we impose additional
average value constraints.

\subsection{Examples}

In this section, we discuss some applications to illustrate our
method.

\begin{itemize}
\item Qubit

We considered the simplest quantum system: one qubit. Since the
state space of a qubit is homotopic to a sphere, this example gives
a graphical representation of how this method works.

Suppose that we want to determine an unknown state $\hat{\rho}$ of a
qubit system, knowing that it is invariant under rotations along the
$\hat{z}$ axis. The generator of the group of rotations along the
$\hat{z}$ axis is $\hat{J}_{z}= \frac{\hbar}{2} \hat{\sigma}_{z}$,
with $\hat{\sigma}_z$ the Pauli matrix given by
\begin{equation}
\hat{\sigma}_z = \begin{pmatrix} 1 & 0 \\ 0 & -1 \end{pmatrix}.
\end{equation}
Therefore, we have to find a state satisfying the condition
\begin{equation}
e^{-\frac{i}{\hbar}\hat{J}_{z}\theta}\hat{\rho}
e^{\frac{i}{\hbar}\hat{J}_{z}\theta}=\hat{\rho}, ~~~ \forall \,
\theta \in [0,2\pi],
\end{equation}
or equivalently, using equation \eqref{solucion 1},
\begin{equation}
\mbox{Tr}(\rho[i\hat{\sigma}_{z},\hat{O}_{j}])=0,  ~~~ \forall
\hat{O_j} \in \{\hat{O_j}\}_{1\leq j\leq 4},
\end{equation}
with $\{\hat{O_j}\}_{1\leq j\leq 4}$ a basis of the space complex
matrices $\mathbb{C}^{2\times 2}$. If we choose
$\hat{O}_{1}=\hat{I}$, $\hat{O}_{2}=\hat{\sigma}_{x}$,
$\hat{O}_{3}=\hat{\sigma}_{y}$ and $\hat{O}_{4}=\hat{\sigma}_{z}$,
the only non-trivial commutators are the following:
\begin{equation}
[i\hat{\sigma}_{z}, \hat{\sigma}_{x}]= -2 \hat{\sigma}_{y}, ~~~~~
[i\hat{\sigma}_{z}, \hat{\sigma}_{y}]= 2 \hat{\sigma}_{x}.
\end{equation}
Therefore, we have two symmetry constraint equations,
\begin{equation}
\langle \hat{\sigma}_{x}\rangle=
\mbox{Tr}(\hat{\rho}\hat{\sigma}_{x})=0, ~~~~ \langle
\hat{\sigma}_{y}\rangle= \mbox{Tr}(\hat{\rho}\hat{\sigma}_{y})=0.
\end{equation}
From the general solution given in equation \eqref{general
solution}, we obtain the density operator which maximizes the
entropy and satisfies the symmetry constraints,
\begin{equation}\label{qbit state}
\hat{\rho} = \frac{e^ {\gamma_{x}\hat{\sigma}_{x}+
\gamma_{y}\hat{\sigma}_{y} }}{Z}, ~~~ Z= \mbox{Tr}(e^
{\gamma_{x}\hat{\sigma}_{x}+ \gamma_{y}\hat{\sigma}_{y}}),
\end{equation}
and, according with equation \eqref{lagrangian multiplier 2}, the
Lagrange multipliers are given by the relations
\begin{equation}
0 = \frac{\partial}{\partial\gamma_{x}}\ln Z, ~~~ 0 =
\frac{\partial}{\partial\gamma_{y}}\ln Z.
\end{equation}
or equivalently,
\begin{equation}\label{multiplicadores}
0 = \frac{\partial Z}{\partial\gamma_{x}}, ~~~ 0 = \frac{\partial
Z}{\partial\gamma_{y}}.
\end{equation}

In order to calculate the explicit expression of the quantum MaxEnt
state, we use the following relation (see equation (2.231) of
\cite{Nielsen})
\begin{equation}\label{Nielsen1}
 e^ {\gamma_{x}\hat{\sigma}_{x}+ \gamma_{y}\hat{\sigma}_{y}} = \frac { e^{\gamma }+ e^{-\gamma} }{2} \hat{I} +\frac { e^{\gamma}- e^{-\gamma}}{2 \gamma} \left( \gamma_{x}\hat{\sigma}_{x}+ \gamma_{y}\hat{\sigma}_{y}\right), ~~~ \gamma = \sqrt{\gamma_x^2 +\gamma_y^2}.
\end{equation}
The partition function is given by $Z=   e^{\gamma }+ e^{-\gamma}$,
and equations \eqref{multiplicadores} take the form
\begin{equation}\label{qbit multipliers}
\frac{\gamma_{x}}{\gamma}\left( e^{\gamma} - e^{-\gamma} \right) = 0
, ~~~ \frac{\gamma_{y}}{\gamma}  \left( e^{\gamma} - e^{-\gamma}
\right) = 0.
\end{equation}
Finally, replacing expressions \eqref{Nielsen1}, \eqref{qbit
multipliers} and the partition function expression into the quantum
MaxEnt \eqref{qbit state}, we obtain $\hat{\rho} =
\frac{\hat{I}}{2}$.

This solution is in agrement with what it was expected: the states
in the Bloch sphere which are invariant under our symmetry are
situated along the zeta axis, and the MaxEnt state is situated in
the center of the sphere.

Now, we can see how this result is modified if we add an extra
condition for the mean value of some observable, for example
$\langle\hat{\sigma}_{z}\rangle=\mbox{Tr}(\hat{\rho}\hat{\sigma}_{z})=a$,
with $-1<a<1$.

From the general solution given in equations \eqref{general
solution}, \eqref{lagrangian multiplier 1} and \eqref{lagrangian
multiplier 2}, we obtain
\begin{equation}\label{qbit state2}
\hat{\rho} = \frac{e^ {\lambda \hat{\sigma}_{z}+
\gamma_{x}\hat{\sigma}_{x}+ \gamma_{y}\hat{\sigma}_{y} }}{Z}, ~~~ Z=
\mbox{Tr}(e^ {\lambda \hat{\sigma}_{z} +\gamma_{x}\hat{\sigma}_{x}+
\gamma_{y}\hat{\sigma}_{y}}),
\end{equation}
\begin{equation}\label{multiplicadores2}
a = \frac{\partial}{\partial\lambda} \ln Z, ~~~ 0 =
\frac{\partial}{\partial\gamma_{x}}\ln Z, ~~~ 0 =
\frac{\partial}{\partial\gamma_{y}}\ln Z.
\end{equation}

Again, we use the relation
\begin{equation}\label{Nielsen2}
e^ { \gamma_{x}\hat{\sigma}_{x}+ \gamma_{y}\hat{\sigma}_{y}+ \lambda
\hat{\sigma}_{z}} = \frac { e^{\gamma }+ e^{-\gamma} }{2} \hat{I}
+\frac { e^{\gamma}- e^{-\gamma}}{2 \gamma} \left(
\gamma_{x}\hat{\sigma}_{x}+ \gamma_{y}\hat{\sigma}_{y}+\lambda
\hat{\sigma}_{z}\right),
\end{equation}
with  $\gamma = \sqrt{\gamma_x^2 +\gamma_y^2+ \lambda^2}$.

The partition function is given by $Z=   e^{\gamma }+ e^{-\gamma}$,
and equations \eqref{multiplicadores2} take the form
\begin{equation}\label{qbit multipliers2}
\frac{\lambda}{\gamma} \left( e^{\gamma} - e^{-\gamma} \right) = a ,
~~~   \frac{\gamma_{x}}{\gamma} \left( e^{\gamma} - e^{-\gamma}
\right) = 0 , ~~~ \frac{\gamma_{y}}{\gamma}  \left( e^{\gamma} -
e^{-\gamma} \right) = 0.
\end{equation}
Finally, replacing expressions \eqref{Nielsen2}, \eqref{qbit
multipliers2} and the partition function expression into the quantum
MaxEnt state \eqref{qbit state2}, we obtain $\hat{\rho} =
\frac{1}{2}\left( \hat{I} + a \hat{\sigma_z}\right)$.

This result is in agrement with what one would expect by appealing
to a geometrical argument. It should be stressed that in the last
case, the maximization of entropy is unnecessary, because there is
only one possible state compatible with the constraints.

\item Qutrit

The second example is a three dimensional quantum system. We want to
estimate the state of the system, knowing that it is invariant under
rotations along the $\hat{z}$ axis. The generator of the rotations
group along the $\hat{z}$ axis is $\hat{J}_{z}$,
\begin{equation}
\hat{J}_z = \hbar \begin{pmatrix} 1 & 0 & 0 \\ 0 & 0 & 0 \\ 0 & 0&
-1\end{pmatrix}.
\end{equation}
Therefore, we have to find a state satisfying the condition
\begin{equation}
e^{-\frac{i}{\hbar}\hat{J}_{z}\theta}\hat{\rho}
e^{\frac{i}{\hbar}\hat{J}_{z}\theta}=\hat{\rho}, ~~~ \forall \,
\theta \in [0,2\pi],
\end{equation}
or equivalently, using our method
\begin{equation}
\mbox{Tr}(\hat{\rho}[i\hat{J}_{z},\hat{O}_{j}])=0,  ~~~ \forall
\hat{O_j} \in \{\hat{O_j}\}_{1\leq j\leq 9 },
\end{equation}
with $\{\hat{O_j}\}_{1\leq j\leq 9}$ a basis of the space of
matrices $\mathbb{C}^{3\times 3}$. We choose the basis given by the
identity $\hat{I}$ and the Gell-Mann matrices $\hat{\lambda}_i$ ($i=
1 \ldots 8$),
\begin{align}
\hat{\lambda}_1 &= \begin{pmatrix} 0 & 1 & 0 \\ 1 & 0 & 0 \\ 0 & 0 & 0 \end{pmatrix}  &&\hat{\lambda}_2 = \begin{pmatrix} 0 & -i & 0 \\ i & 0 & 0 \\ 0 & 0 & 0 \end{pmatrix}   &&\hat{\lambda}_3 &= \begin{pmatrix} 1 & 0 & 0 \\ 0 & -1 & 0 \\ 0 & 0 & 0 \end{pmatrix} &&\hat{\lambda}_4 = \begin{pmatrix} 0 & 0 & 1 \\ 0 & 0 & 0 \\ 1 & 0 & 0 \end{pmatrix} \nonumber\\
 \hat{\lambda}_5 &= \begin{pmatrix} 0 & 0 & -i \\ 0 & 0 & 0 \\ i & 0 & 0 \end{pmatrix} &&\hat{\lambda}_6 = \begin{pmatrix} 0 & 0 & 0 \\ 0 & 0 & 1 \\ 0 & 1 & 0 \end{pmatrix}   &&\hat{\lambda}_7 &= \begin{pmatrix} 0 & 0 & 0 \\ 0 & 0 & -i \\ 0 & i & 0 \end{pmatrix} &&\hat{\lambda}_8 =  \begin{pmatrix} \frac{1}{\sqrt{3}} & 0 & 0 \\ 0 & \frac{1}{\sqrt{3}} & 0 \\ 0 & 0 & \frac{-2}{\sqrt{3}} \end{pmatrix} \nonumber.
\end{align}
In this case, the only non-trivial commutators are the following:
\begin{align}
[i\hat{J}_{z}, \hat{\lambda}_{1}]&= -\hbar  \hat{\lambda}_{2},
&&[i\hat{J}_{z}, \hat{\lambda}_{2}]= \hbar  \hat{\lambda}_{1},
&&[i\hat{J}_{z}, \hat{\lambda}_{4}]= -2 \hbar  \hat{\lambda}_{5},
\nonumber\\  [i\hat{J}_{z}, \hat{\lambda}_{5}]&= 2 \hbar
\hat{\lambda}_{4}, &&[i\hat{J}_{z}, \hat{\lambda}_{6}]= -\hbar
\hat{\lambda}_{7},    &&[i\hat{J}_{z}, \hat{\lambda}_{7}]= \hbar
\hat{\lambda}_{6}.
\end{align}
Therefore, we have six symmetry constraint equations,
\begin{equation}
\mbox{Tr}(\hat{\rho}\hat{\lambda}_i)=0, ~~~ \mbox{for}~ i= 1, \,
2,\, 4,\, 5,\, 6,\, 7.
\end{equation}
From the general solution given in equation \eqref{general
solution}, we obtain the density operator which maximizes the
entropy and satisfies the symmetry constraints,
\begin{equation}\label{qtrit state}
\hat{\rho} = \frac{e^ {\hat{M}}}{Z}, ~~~ Z= \mbox{Tr}(e^ {\hat{M}}),
\end{equation}
with
\begin{equation}
\hat{M} =
\gamma_{1}\hat{\lambda}_1+\gamma_{2}\hat{\lambda}_2+\gamma_{4}\hat{\lambda}_4+
\gamma_{5}\hat{\lambda}_5+ \gamma_{6}\hat{\lambda}_6+
\gamma_{7}\hat{\lambda}_7 = \begin{pmatrix} 0 & \gamma_1- i \gamma_2
& \gamma_4- i \gamma_5 \\ \gamma_1+ i \gamma_2 & 0 & \gamma_6- i
\gamma_7 \\ \gamma_4+ i \gamma_5 & \gamma_6 + i \gamma_7 & 0
\end{pmatrix},
\end{equation}
and the Lagrange multipliers are given by
\begin{equation}\label{qtritmultiplicadores}
0 = \frac{\partial}{\partial\gamma_{i}}\ln Z, ~~~ \mbox{for}~ i= 1,
\, 2,\, 4,\, 5,\, 6,\, 7.
\end{equation}

Since $\hat{M}$ is an Hermitian matrix, it is diagonalizable. Then,
there exists a unitary matrix $\hat{U} \in \mathbb{C}^{3\times 3} $
and a real diagonal matrix $\hat{D} \in \mathbb{C}^{3\times 3}$,
\begin{equation}
\hat{D} = \begin{pmatrix} d_1 & 0 & 0 \\ 0 & d_2 & 0 \\ 0 & 0 & d_3
\end{pmatrix}
\end{equation}
where $d_1$, $d_2$, $d_3$ are the eigenvalues of $\hat{M}$, such
that $\hat{M} = \hat{U}\hat{D}\hat{U^{-1}}$.

Therefore, $Z= \mbox{Tr}(e^ {\hat{M}})= e^ {d_1}+e^{d_2}+ e^{d_3}$,
and the equations \eqref{qtritmultiplicadores} take the form

\begin{equation}
0 =  e^ {d_1} \frac{\partial d_1}{\partial\gamma_{i}}+e^ {d_2}
\frac{\partial d_2}{\partial\gamma_{i}}+e^ {d_3} \frac{\partial
d_3}{\partial\gamma_{i}} , ~~~ \mbox{for}~ i= 1, \, 2,\, 4,\, 5,\,
6,\, 7.
\end{equation}

The explicit expression of the eigenvalues in terms of the Lagrange
multipliers and their derivatives are cumbersome. However, it can be
shown that the solution for the Lagrange multipliers is
$\gamma_{1}=\gamma_{2}= \gamma_{4}=  \gamma_{5} =\gamma_{6}=
\gamma_{7}= 0$. Therefore, the quantum MaxEnt state is $\hat{\rho} =
\frac{\hat{I}}{3}$.

Now, we can see how this result is modified if we add an extra
condition for the mean value of some observable, for example
$\langle\hat{J}_{z}\rangle=\mbox{Tr}(\hat{\rho}\hat{J}_{z})=\hbar
a$, with $a \in \mathbb{R} $. In this case, we have seven constraint
equations
\begin{equation}
\mbox{Tr}(\hat{\rho}\hat{\lambda}_i)=0, ~~~ \mbox{for}~ i= 1, \,
2,\, 4,\, 5,\, 6,\, 7, ~~~~ \mbox{Tr}\left(
\hat{\rho}\frac{\hat{J}_z}{\hbar}\right) =a.
\end{equation}
Again, the general solution is given by
\begin{equation}\label{qtrit state2}
\hat{\rho} = \frac{e^ {\hat{M}}}{Z}, ~~~ Z= \mbox{Tr}(e^ {\hat{M}}),
\end{equation}
with $ \hat{M}
=\gamma_{0}\frac{\hat{J}_z}{\hbar}+\gamma_{1}\hat{\lambda}_1+\gamma_{2}\hat{\lambda}_2+\gamma_{4}\hat{\lambda}_4+
\gamma_{5}\hat{\lambda}_5+ \gamma_{6}\hat{\lambda}_6+
\gamma_{7}\hat{\lambda}_7$, and the Lagrange multipliers are given
by
\begin{equation}
a = \frac{\partial}{\partial\gamma_{0}} \ln Z, ~~~~ 0 =
\frac{\partial}{\partial\gamma_{i}}\ln Z, ~~~ \mbox{for}~ i= 1, \,
2,\, 4,\, 5,\, 6,\, 7.
\end{equation}

In order to solve this problem, we use a numerical algorithm. We
consider two cases, $a = -1/2, \, \, 1/2$, and we minimize the
following expresion:
\begin{equation}\nonumber
\Delta = \left[ \mbox{Tr}\left(
\hat{\rho}\frac{\hat{J}_z}{\hbar}\right) - a\right]^2 + \sum_{i\in
\{1,2,4,5,6,7\}}\left[ \mbox{Tr}(\hat{\rho}\hat{\lambda}_i)\right]^2
,
\end{equation}
taking into account the general form of $\hat{\rho}$ given in
equation \eqref{qtrit state2}. The results obtained are the
following:

~~~~ \textit{For $a=-1/2$:}
    \begin{equation}
    \hat{ \rho} = \begin{pmatrix} 0.116 & 0 & 0 \\ 0 & 0.268 & 0 \\ 0 & 0 & 0.616 \end{pmatrix}
    \end{equation}

~~~~ \textit{For $a=1/2$:}
    \begin{equation}
    \hat{ \rho} = \begin{pmatrix} 0.616 & 0 & 0 \\ 0 & 0.268 & 0 \\ 0 & 0 & 0.116 \end{pmatrix}
    \end{equation}

\noindent The above examples show that maximizing state departs from
the maximally mixed state, depending on the conditions imposed on
the mean values of the observables.

\item Two qubits

The third example is a composite system of two spin $1/2$ systems.
We want to estimate the state the system, knowing that the first
spin is invariant under rotations along the $\hat{z}$ axis. The
generator of the symmetry is $\hat{J}_z \otimes \hat{I}$. Moreover,
we consider an extra condition for the mean value of the $\hat {x}$
component of the total spin of the system, i.e.
$\hat{J}_{1x}+\hat{J}_{2x}$,
\begin{equation}
\langle\hat{J}_{1x}+\hat{J}_{2x}\rangle=\mbox{Tr}\left[
\hat{\rho}\left( \hat{J}_{1x}+\hat{J}_{2x}\right) \right] =\hbar
a/2.
\end{equation}

Therefore, we have to find a state satisfying the conditions
\begin{align}
&\mbox{Tr}(\hat{\rho}[i\hat{J}_z \otimes \hat{I},\hat{O}_{j}])=0,
~~~ \forall \hat{O_j}
\in \{\hat{O_j}\}_{1\leq j\leq 16 }, \\
&\mbox{Tr}\left[ \hat{\rho}\left(
\hat{\sigma}_{1x}+\hat{\sigma}_{2x}\right) \right] = a.
\end{align}
with $\{\hat{O_j}\}_{1\leq j\leq 16}$ a basis of the space of
matrices $\mathbb{C}^{4\times 4}$. If we choose the following basis
\begin{align}
\hat{O}_1 &= \begin{pmatrix} 1 & 0 & 0 & 0 \\ 0 & 0 & 0 & 0 \\ 0 & 0
& 0 & 0 \\ 0 & 0 & 0 & 0 \end{pmatrix}  &&\hat{O}_2 =
\begin{pmatrix} 0 & 0 & 0 & 0 \\ 0 & 1 & 0 & 0 \\ 0 & 0 & 0 & 0 \\ 0
& 0 & 0 & 0 \end{pmatrix}  &&
\hat{O}_3 &= \begin{pmatrix}  0 & 0 & 0 & 0 \\ 0 & 0 & 0 & 0 \\ 0 & 0 & 1 & 0 \\ 0 & 0 & 0 & 0\end{pmatrix} &&\hat{O}_4 = \begin{pmatrix}  0 & 0 & 0 & 0 \\ 0 & 0 & 0 & 0 \\ 0 & 0 & 0 & 0 \\ 0 & 0 & 0 & 1 \end{pmatrix} \nonumber\\
\hat{O}_5 &= \begin{pmatrix}  0 & 1 & 0 & 0  \\  1 & 0 & 0 & 0  \\
0 & 0 & 0 & 0 \\  0 & 0 & 0 & 0 \end{pmatrix} &&\hat{O}_6 =
\begin{pmatrix}  0 & i & 0 & 0  \\  -i & 0 & 0 & 0 \\ 0 & 0 & 0 & 0
\\  0 & 0 & 0 & 0  \end{pmatrix}  && \hat{E}_7 &= \begin{pmatrix}
0 & 0 & 1 & 0 \\  0 & 0 & 0 & 0  \\   1 & 0 & 0 & 0 \\ 0 & 0 & 0 & 0
\end{pmatrix} &&\hat{E}_8 = \begin{pmatrix}   0 & 0 & i & 0 \\ 0 & 0
& 0 & 0  \\  -i & 0 & 0 & 0 \\ 0 & 0 & 0 & 0  \end{pmatrix}
\nonumber
\nonumber\\
\hat{O}_9 &= \begin{pmatrix}  0 & 0 & 0 & 1  \\  0 & 0 & 0 & 0  \\
0 & 0 & 0 & 0 \\  1 & 0 & 0 & 0 \end{pmatrix} &&\hat{O}_{10} =
\begin{pmatrix}  0 & 0 & 0 & i  \\  0 & 0 & 0 & 0 \\ 0 & 0 & 0 & 0
\\  -i & 0 & 0 & 0  \end{pmatrix}  && \hat{E}_{11} &=
\begin{pmatrix}   0 & 0 & 0 & 0 \\  0 & 0 & 1 & 0  \\   0 & 1 & 0 &
0 \\ 0 & 0 & 0 & 0  \end{pmatrix} &&\hat{E}_{12} = \begin{pmatrix}
0 & 0 & 0 & 0 \\ 0 & 0 & i & 0  \\   0 & -i & 0 & 0 \\ 0 & 0 & 0 & 0
\end{pmatrix}
\nonumber\\
\hat{O}_{13} &= \begin{pmatrix}  0 & 0 & 0 & 0  \\  0 & 0 & 0 & 1
\\  0 & 0 & 0 & 0 \\  0 & 1 & 0 & 0 \end{pmatrix} &&\hat{O}_{14} =
\begin{pmatrix}  0 & 0 & 0 & 0  \\  0 & 0 & 0 & i \\ 0 & 0 & 0 & 0
\\  0 & -i & 0 & 0  \end{pmatrix}  && \hat{E}_{15} &=
\begin{pmatrix}   0 & 0 & 0 & 0 \\  0 & 0 & 0 & 0  \\   0 & 0 & 0 &
1 \\ 0 & 0 & 1 & 0  \end{pmatrix} &&\hat{E}_{16} = \begin{pmatrix}
0 & 0 & 0 & 0 \\ 0 & 0 & 0 & 0  \\   0 & 0 & 0 & i \\ 0 & 0 & -i & 0
\end{pmatrix} \nonumber
\end{align}
we obtain the following equations
\begin{equation}\label{constraints quatrit}
\mbox{Tr}(\hat{\rho}\hat{E}_i)=0, ~~~ \mbox{for}~ i= 1, \ldots, 8,
~~~~~ \mbox{Tr}\left[ \hat{\rho}\left(
\hat{\sigma}_{1z}+\hat{\sigma}_{2z}\right) \right] = a,
\end{equation}
with the matrixes $\hat{E}_i$ ($i= 1, \ldots, 8$) given by
\begin{align}
\hat{E}_1 &= \begin{pmatrix} 0 & 0 & -i & 0 \\ 0 & 0 & 0 & 0 \\ i &
0 & 0 & 0 \\ 0 & 0 & 0 & 0 \end{pmatrix}  &&\hat{E}_2 =
\begin{pmatrix} 0 & 0 & 1 & 0 \\ 0 & 0 & 0 & 0 \\ 1 & 0 & 0 & 0 \\ 0
& 0 & 0 & 0 \end{pmatrix}  &&
\hat{E}_3 &= \begin{pmatrix}  0 & 0 & 0 & -i \\ 0 & 0 & 0 & 0 \\ 0 & 0 & 0 & 0 \\ i & 0 & 0 & 0\end{pmatrix} &&\hat{E}_4 = \begin{pmatrix}  0 & 0 & 0 & 1 \\ 0 & 0 & 0 & 0 \\ 0 & 0 & 0 & 0 \\ 1 & 0 & 0 & 0 \end{pmatrix} \nonumber\\
\hat{E}_5 &= \begin{pmatrix}  0 & 0 & 0 & 0  \\  0 & 0 & -i & 0  \\
0 & i & 0 & 0 \\  0 & 0 & 0 & 0 \end{pmatrix} &&\hat{E}_6 =
\begin{pmatrix}  0 & 0 & 0 & 0  \\  0 & 0 & 1 & 0 \\ 0 & 1 & 0 & 0
\\  0 & 0 & 0 & 0  \end{pmatrix}  && \hat{E}_7 &= \begin{pmatrix}
0 & 0 & 0 & 0 \\  0 & 0 & 0 & -i  \\   0 & 0 & 0 & 0 \\ 0 & i & 0 &
0  \end{pmatrix} &&\hat{E}_8 = \begin{pmatrix}   0 & 0 & 0 & 0 \\ 0
& 0 & 0 & 1  \\   0 & 0 & 0 & 0 \\ 0 & 1 & 0 & 0  \end{pmatrix}
\nonumber.
\end{align}

From the general solution given in equation \eqref{general
solution}, we obtain the density operator which maximizes the
entropy and satisfies the constraint equations \eqref{constraints
quatrit},
\begin{equation}\label{quatrit state}
\hat{\rho} = \frac{e^ {\hat{M}}}{Z}, ~~~ Z= \mbox{Tr}(e^ {\hat{M}}),
\end{equation}
with $\hat{M} = \gamma_{0} \left(
\hat{\sigma}_{1x}+\hat{\sigma}_{2x}\right) +\sum_{i=1}^{8}
\gamma_{i}\hat{E}_i$, and the Lagrange multipliers are given by
\begin{equation}\label{twoqubitsmultiplicadores}
a = \frac{\partial}{\partial\gamma_{0}}\ln Z,  ~~~~ 0 =
\frac{\partial}{\partial\gamma_{i}}\ln Z, ~~~ \mbox{for}~ i= 1,
\ldots , 8.
\end{equation}

In order to solve this problem, we again use a numerical algorithm.
We consider three cases, $a = -1/2, \, 0, \, 1/2$, and we minimize
the following expresion:
\begin{equation}\nonumber
\Delta = \left[ \mbox{Tr}\left( \hat{\rho}\left(
\hat{\sigma}_{1x}+\hat{\sigma}_{2x}\right) \right) - a\right]^2
+\sum_{i=1}^{8}\left[ \mbox{Tr}(\hat{\rho}\hat{E}_1)\right]^2
\end{equation}
taking into account the general form of $\hat{\rho}$ given in
equation \eqref{quatrit state}.

The results obtained are the following:

~~~~ \textit{For $a=-1$:}
\begin{equation}
\hat{ \rho} = \begin{pmatrix} 0.25 & -0.25 & 0 & 0\\ -0.25 & 0.25 &
0& 0 \\ 0 & 0 & 0.25 & -0.25 \\ 0 & 0 & -0.25 & 0.25 \end{pmatrix}
\end{equation}

~~~~ \textit{For $a=0$:}
\begin{equation}
\hat{ \rho} = \begin{pmatrix} 0.25 & 0 & 0 & 0\\ 0 & 0.25 & 0& 0 \\
0 & 0 & 0.25 & 0 \\ 0 & 0 & 0 & 0.25 \end{pmatrix}
\end{equation}

~~~~ \textit{For $a=1$:}
\begin{equation}
\hat{ \rho} = \begin{pmatrix} 0.25 & 0.25 & 0 & 0\\ 0.25 & 0.25 & 0&
0 \\ 0 & 0 & 0.25 & 0.25 \\ 0 & 0 & 0.25 & 0.25 \end{pmatrix}
\end{equation}

\noindent This example shows that the group structure is richer as
the number of particles grows, as is expressed by the action of the
local groups.

\end{itemize}

\section{Conclusions}\label{s:Conclusions}

In this work, we continued studying the problem posed in
\cite{Holik-Massri-Plastino-IJGMMP-2016}, namely, the state
estimation of probabilistic models with symmetries represented by
groups of transformations. A complete tomography of a multi-particle
state requires a number of measurements that grows exponentially
with the number of particles. This number can be significantly
reduced whenever some a priori information is available. Our
approach can be useful for quantum state estimation problems in
which this information is given in terms of symmetries of the
system.

First, we revised the traditional version of the classical and
quantum MaxEnt estimation problem, and we reformulated them
including additional symmetry constraints. We presented the
classical MaxEnt estimation problem for finite systems with symmetry
constraints. Then, we considered the quantum MaxEnt problem for
systems with finite dimension and with symmetries represented by Lie
and finite groups.

Finally, we proved that the symmetry constraints can be restated as
a set of linear equations, and we presented a solution. We
illustrated how our method works by showing some examples.

\section*{Acknowledgments}

This research was funded by the Consejo Nacional de Investigaciones
Cient\'ificas y T\'ecnicas (CONICET).


\end{document}